\documentclass[12pt]{article}
\usepackage[utf8]{inputenc}
\usepackage[T1]{fontenc}
\usepackage{amsmath,amssymb,graphicx,natbib,geometry,hyperref}
\title{An Identifiability Audit of One-Parameter Structural Corrections to the Radial Acceleration Relation in SPARC}
\author{Lukas A. Sosna\\ \small Independent Researcher \\ \small ORCID: 0009-0004-1266-9729 \\ \small LukasASosna@gmail.com}
\date{}
\begin{document}
\maketitle
\begin{abstract}
We ask whether any one-parameter structural correction to the radial acceleration relation (RAR) can be uniquely recovered from SPARC rotation curves, and answer with an identifiability audit: each candidate is benchmarked against per-galaxy nuisance freedom, with predictive scoring against mass-only and data-quality baselines. In the full sample (N = 126, fixed mass-to-light protocol) the answer is no: a hybrid compactness term improves the fit, but zero-point freedom absorbs the gain, and in cross-validation the model fails to out-predict a mass-only baseline (MSPE ratio 0.99) and loses to a quality-flag baseline (1.22). One regime retains structural information: in gas-dominated, low-acceleration disks --- where MOND's strict locality and $\Lambda$CDM feedback models diverge most sharply --- the RAR residual correlates with compactness (r = 0.46, p = 1.3 $\times 10^{-4}$), remains significant under hierarchical partial pooling ($\beta$ = 0.23, p = 1.7 $\times 10^{-5}$; N = 63 galaxies), and survives a canonical joint control for quality, sampling, mass, inclination error, and first-order pressure support (partial r = 0.30, p = 0.02), with all significant results passing a Benjamini--Hochberg correction over the declared 27-test family. Three limits temper that survival: the correlation is not significant under rank-based control over the widest proxy set; it resides in the faint dwarfs independent surveys do not reach; and after mass control it is shared across the mass--size manifold rather than attributable to compactness. Its multiplicity-corrected margin is narrow --- a factor of 1.4 above threshold. Pressure support brackets the interpretation --- an isotropic drift correction absorbs about a quarter of the amplitude, while a Jeans treatment overcorrects resolved test cases --- so we cannot tell whether the residual is structural physics or unmodelled kinematics. The audit's product is the extraction limit: any claimed structural correction must clear the 0.106 dex per-galaxy nuisance floor (0.080 dex of absorbable scatter against the structural term's 0.015), a mass-only predictive baseline, and data-quality stratification before a physical reading is warranted.
\end{abstract}

\section{Introduction}

Several groups have reported weak correlations between RAR residuals and galaxy properties (\citealp{Lelli2017}; \citealp{Rodrigues2018}; \citealp{Li2018}; \citealp{Marra2020}). What has not been done is to test whether any such correlation is uniquely recoverable: whether a one-parameter structural correction to the RAR can be separated from the galaxy-level zero-point freedom that distance, inclination, and mass-to-light uncertainties already permit. That test is the subject of this paper. Per-galaxy nuisance parameters are themselves standard in SPARC rotation-curve fitting (\citealp{Li2018}; \citealp{Desmond2023}); what has not been done is to use that freedom as an explicit identifiability benchmark for a proposed structural correction, and to score the correction predictively against mass-only and data-quality baselines. We call the result an identifiability audit because the question is not whether a residual exists but whether the data can attribute it.

Under Modified Newtonian Dynamics (MOND; \citealp{Milgrom1983}; \citealp{Famaey2012}), the RAR is exact: observed acceleration is a universal function of baryonic acceleration, with no room for a secondary structural term. Under $\Lambda$CDM, baryonic feedback can sculpt dark-matter halo profiles in structure-dependent ways (\citealp{DiCintio2014}; \citealp{Keller2016}), and the predicted imprints are strongest in gas-dominated systems where halo response and feedback efficiency vary most. A structural term that survived rigorous nuisance control would be evidence against the strict MOND prediction. A structural term that dissolved into per-galaxy systematics would show that claims of this kind, several of which exist in the literature, do not yet clear the evidential bar.

That evidential bar has become concrete because the observational regime of interest has narrowed. Both strict-RAR MOND and structure-dependent $\Lambda$CDM feedback models predict their largest divergences in the same corner of the disk-galaxy parameter space: gas-dominated, low-acceleration systems, where the baryonic mass budget is measured directly and the halo-response predictions are strongest. SPARC contains roughly sixty such systems. Whether a residual in this regime is physical or systematic is not a question the field has resolved, and it is the question this paper's audit is designed to sharpen rather than settle.

Detection is limited by galaxy-level systematics before it is limited by statistics. In SPARC, under a fixed mass-to-light protocol, distance uncertainties (~0.08 dex), inclination errors (~0.06 dex), and mass-to-light scatter combine to roughly 0.10 dex per galaxy. Any one-parameter correction with amplitude at or below this floor cannot be uniquely separated from observational noise in the full sample. This floor is specific to SPARC under the fixed protocol adopted here; hierarchical analyses with different modelling choices report lower intrinsic scatter (\citealp{Desmond2023} finds $\sigma_{\rm int}$ $\approx$ 0.034 dex), and independent surveys are beginning to probe the relation with different systematics (\citealp{Varasteanu2025} measure $\sigma_{\rm int}$ $\approx$ 0.045 dex in MIGHTEE-HI). The extraction boundary will move as data improve. Our purpose is to measure where it sits now, and what gets through it.

The regime structure of the problem matters. In stellar-dominated galaxies the baryonic mass budget depends on an assumed mass-to-light ratio, which adds a systematic layer. In gas-dominated galaxies the baryonic mass is measured directly from 21 cm flux, which removes that layer but exposes another: kinematic extraction in gas-rich dwarfs must contend with beam smearing, pressure support, and non-circular motions (\citealp{deBlok1997}; \citealp{Oh2015}; \citealp{Iorio2017}; \citealp{Oman2019}). The gas-dominated regime has the cleanest baryonic masses and the most dangerous kinematics. That tension is central to everything that follows.

A further complication comes from the mass-modelling protocol itself. When the stellar mass-to-light ratio is tuned per galaxy to minimize rotation-curve residuals, the resulting baryonic mass absorbs kinematic information, inducing covariance between structural quantities and dynamical residuals even without physical coupling. The protocol is therefore a confound to be controlled explicitly (Section 3.4).

We use compactness, $\lambda = GM_{\rm bar}/(R_{\rm eff}c^{2})$, as the entry variable for the audit. It is the lowest-order dimensionless potential constructible from the two observables available for every SPARC galaxy, and it condenses the mass--size plane into one scalar suitable for a controlled one-parameter test. Nothing here privileges compactness as fundamental. Any scalar built from mass and size correlates with the same structural manifold, and Sections 3.10 and 3.11 test directly whether the results are specific to compactness or shared across the manifold. We already note the answer: shared. Because $\lambda$ is constructed from $M_{\rm bar}$, it correlates strongly with mass by construction (r = 0.956, Section 3.1); every inference in this paper that bears physical weight is therefore made after mass control or against mass-based baselines. Readers should treat $\lambda$ as a probe of the mass--size manifold, not as a candidate fundamental variable.

A note on language. What we observe are residuals: statistical departures from the RAR baseline. Whether a residual reflects a stable coupling to galaxy structure, and whether any such coupling survives nuisance control and coordinate choice, are separate questions, and the audit is built to keep them separate. In the full sample the answer to the second question is no. In the gas-dominated regime it is unresolved, and we say so rather than choosing.

The paper delivers four results. First, a quantified extraction limit: no one-parameter structural correction clears per-galaxy nuisance freedom in the full SPARC sample, and the compactness model has no predictive advantage over a mass-only baseline under cross-validation. Second, a localized residual in gas-dominated, low-acceleration disks that survives every internal control we can construct, including a canonical joint quality-and-kinematics control specified in advance of computation, but whose physical origin the data cannot settle. Third, a demonstration that the residual is a property of the mass--size manifold rather than of any single structural coordinate. Fourth, four methodological confounds, including data-quality entanglement, that apply to any structure--dynamics residual analysis in resolved rotation-curve samples. The analysis is restricted to the SPARC field-disk regime and to one-parameter structural models.

\section{Data and Methods}

\subsection{Data source and sample definition}

We use the SPARC database of nearby disk galaxies (\citealp{Lelli2016}), including the photometric, structural, and rotation-curve source tables together with the associated mass-model products. The frozen sample applies a quality cut of Q $\leq$ 3, an inclination cut of $30^{\circ} < i < 80^{\circ}$, and a requirement of at least five valid rotation-curve points per galaxy. The resulting sample contains 126 galaxies and 2,709 valid rotation-curve points. The quality distribution is Q = 1: 74 galaxies; Q = 2: 44; Q = 3: 8. Because most published SPARC analyses adopt a stricter i > 40$^{\circ}$ cut (\citealp{Lelli2017}; \citealp{Li2018}), all headline results are re-verified at that threshold in Section 3.7; the localized signal strengthens there, so the wider cut is conservative for our conclusions.

The effective radius $R_{\rm eff}$ is used consistently across all protocol comparisons. This symmetric radius policy avoids importing operator asymmetry into the analysis: mixed-operator comparisons ($R_{\rm eff}$ for one protocol, $R_{\rm disk}$ for another) can amplify apparent protocol differences through geometry alone, and the operator non-invariance between $R_{\rm eff}$ and $R_{\rm disk}$ introduces roughly 0.176 dex of scatter. Under both operators the fixed-protocol structure--residual correlation is nearly identical (r = 0.326 vs r = 0.323), so within-protocol results are stable against this choice.

\subsection{Compactness and baryonic mass}

We define compactness as

\begin{equation*}
\lambda = \frac{G M_{\rm bar}}{R_{\rm eff}\, c^{2}}
\end{equation*}

with

\begin{equation*}
M_{\rm bar} = \Upsilon_{*}\, L_{[3.6]} + 1.33\, M_{\rm HI}
\end{equation*}

where the factor 1.33 accounts for cosmological helium, following the SPARC convention (\citealp{Lelli2016}); some recent works adopt 1.36, a difference that is negligible at the precision of this analysis. For the structural quantity $\lambda$ we adopt $\Upsilon_{*}$ = 0.5 $M_{\odot}/L_{\odot}$ at 3.6 $\mu$m applied to the total luminosity (\citealp{Meidt2014}; \citealp{Schombert2019}). In the rotation-curve decomposition, the fixed protocol applies $\Upsilon_{\rm disk}$ = 0.5 to the disk component and $\Upsilon_{\rm bul}$ = 0.7 to the bulge component, the standard SPARC convention (\citealp{Lelli2016}); bulges are rare in the gas-dominated subsample where the localized signal resides, so this distinction affects only the stellar-dominated regime.

\subsection{Fixed and tuned mass-to-light protocols}

To assess how sensitive any structural signal is to the mass-modelling strategy, we compare two approaches.

\textbf{Protocol A (Fixed):} $\Upsilon_{\rm disk}$ = 0.5, $\Upsilon_{\rm bul}$ = 0.7 applied uniformly. No kinematic information enters the mass normalisation.

\textbf{Protocol B (Tuned):} $\Upsilon_{*}$ is adjusted per galaxy to minimise rotation-curve residuals, following \citet{Li2018}. This absorbs structural information into the mass scale, because the optimisation adjusts the baryonic amplitude to match observed velocities.

The distinction matters because $\lambda$ shares a mass axis with RAR residuals: any protocol that adjusts $M_{\rm bar}$ per galaxy can induce covariance between $\lambda$ and the residuals without physical coupling.

\subsection{Residual definition and statistical conventions}

For each rotation-curve point we compute $g_{\rm bar}$ from the signed velocity decomposition (|V|V convention) and $g_{\rm obs}$ = $V_{\rm obs}^{2}$/R, and define the point residual as $\delta$ = $\log_{10}$($g_{\rm obs}$) - $\log_{10}$($g_{\rm RAR}$($g_{\rm bar}$)), where $g_{\rm RAR}$ is the \citet{McGaugh2016} interpolating function with $a_0$ = 1.2 $\times 10^{-10}$ m s$^{-2}$ held fixed. Galaxy-level correlation analyses use the median point residual per galaxy, for resistance against localized non-circular motions and individual outlier points within a rotation curve; cross-validation (Section 3.3) uses the mean, aligning the target with the mean-squared-prediction-error objective against which the models are scored. Both conventions are applied consistently and stated wherever a number is reported.

Because this paper reports many statistical tests, we declare the test family explicitly: 27 confirmatory tests spanning the protocol comparison, regime split, permutation tests, quality stratification, asymmetric-drift sensitivity, hierarchical models, manifold comparisons, and the canonical joint partial correlation of Section 4.4. We apply a Benjamini--Hochberg correction at FDR = 0.05 across this family. Seventeen of the 27 tests survive; every result reported as significant in this paper is among the survivors, and every test that fails the correction is one we report as null or marginal in any case. No conclusion in this paper depends on a test that fails the correction. Tests are classed confirmatory if they bear on a claim made in the abstract or conclusions; sensitivity sweeps, sub-sample removals, and robustness variants reported in Section 3 are exploratory and are not counted in the family. The full family, with raw and adjusted p-values, is tabulated in the supplementary material.

\subsection{Model framework}

We compare four models using the Bayesian Information Criterion, BIC $= k\ln(n) - 2\ln(\hat{L})$ (\citealp{Kass1995}). BIC is used as a comparative identifiability diagnostic within a fixed Gaussian log-likelihood framework, not as a test of physical truth; where $\Delta$BIC values are very large the direction of the comparison is unlikely to reverse under alternative criteria, but precise magnitudes should not be over-interpreted. Because rotation-curve points within a galaxy share distance, inclination, and mass-to-light ratio, we report the galaxy-compressed comparison (n = 126) as primary and the point-level comparison (n = 2,709) as a diagnostic. To avoid exclusive dependence on BIC we add galaxy-level cross-validation against three baselines (Section 3.3).

\textbf{M0 (the RAR):} $g_{\rm obs} = g_{\rm bar}/[1-\exp(-\sqrt{g_{\rm bar}/a_0})]$ with $a_0$ fixed. k = 1.

\textbf{M1 (global-only modulation):} $g_{\rm obs}$ = A($(\lambda/\lambda_0)^{\gamma}$ $g_{\rm bar}$, with $\lambda_0$ = $10^{-7}$. k = 2. A boundary test; no serious framework proposes local accelerations depend solely on global compactness.

\textbf{M2 (hybrid):} $g_{\rm obs}$ = $g_{\rm RAR}$($g_{\rm bar}$) $\times$ [1 + $\varepsilon$($(\lambda/\lambda_0)^{\gamma}$], with $\varepsilon$ and $\gamma$ free. k = 3.

\textbf{M3 (nuisance intercept):} each galaxy receives a free additive offset $\delta_j$ in log space, no $\lambda$ dependence. k = 126, evaluated at the point level where degrees of freedom remain highly positive. M3 is not a competing physical model and its BIC "win" over M2 is guaranteed by construction whenever per-galaxy zero-point structure exists in the data. Its value is diagnostic: it measures how much per-galaxy variance (0.080 dex, Section 3.3) is available to absorb any one-parameter gain, which is the floor a structural term must clear to be considered identifiable. The properly powered test of whether a structural term survives principled per-galaxy freedom is the hierarchical random-intercept model of Section 4.4, which occupies the space between M2 and M3.

\subsection{Regression and scatter analysis}

The compactness--mass relation is fitted with Orthogonal Distance Regression (scipy.odr), which accounts for measurement errors in both coordinates. For full determinism we declare the error model: $\sigma$ = 0.094 dex assigned to both coordinates, the quadrature sum of the measurement error budget below. Under this model the ODR slope is 0.748 $\pm$ 0.020; OLS gives 0.727 for comparison, and a galaxy-level bootstrap (2,000 resamples, seed 42) yields a standard error of 0.019 and a 95 per cent CI of [0.711, 0.787], so the parametric uncertainty is not optimistic. The measurement error budget includes distance (~0.08 dex, dominant), photometry (~0.004 dex), H I mass (~0.02 dex), and effective radius (~0.04 dex), combining to $\sigma_{\rm meas}$ = 0.094 dex.

\subsection{Pipeline sensitivity verification}

To confirm the fixed-protocol pipeline retains sensitivity to genuine signals, we inject synthetic correlations of known amplitude into the residual data and measure recovery across 200 Monte Carlo realisations per amplitude. The false-positive rate at zero injected signal is 5.0 per cent, matching the nominal $\alpha$ = 0.05. Recovery reaches 93.6 per cent at $\varepsilon$ = 0.04 and 100 per cent at $\varepsilon$ $\geq$ 0.06, well below the observed M2 amplitude (|$\varepsilon$| = 0.077). Null results under the fixed protocol therefore reflect absence of extractable signal, not absence of power. All derived quantities were independently recomputed from the SPARC MRT source tables, including the signed-velocity treatment, $\Upsilon$ scaling from the documented M/L = 1 columns, and the helium-inclusive gas convention; galaxy-level median residuals reproduce the frozen analysis values to machine precision ($\leq10^{-4}$ dex).

\section{Results}

\subsection{Compactness as a structural coordinate}

\begin{figure}[htbp]
\centering
\includegraphics[width=\linewidth]{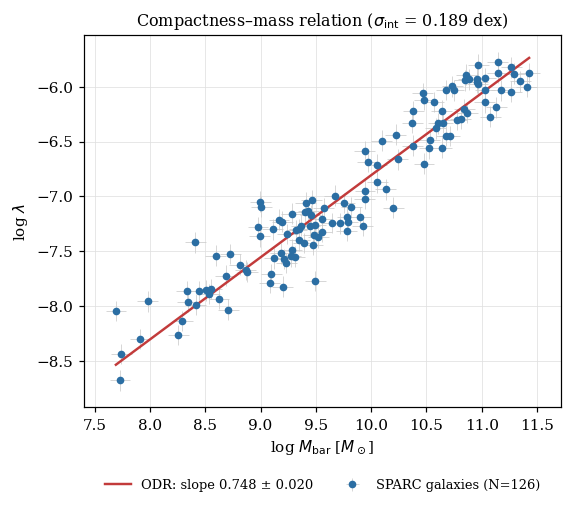}
\caption{Compactness--mass relation with the declared-model ODR fit; error bars show the 0.094 dex measurement budget on both coordinates.}
\label{fig:1}
\end{figure}

The compactness--mass relation is tight (Fig.~\ref{fig:1}). ODR under the declared error model gives

\begin{equation*}
\log_{10}(\lambda) = (0.748 \pm 0.020)\,\log_{10}(M_{\rm bar}) + {\rm const}
\end{equation*}

with Pearson r = 0.956 (p < $10^{-65}$). The slope deviates from virial scaling (s = 1), reflecting the mass--size relation; because log $\lambda$ = log $M_{\rm bar}$ - log $R_{\rm eff}$ + const, the compactness slope is algebraically one minus the mass--size slope, corresponding to $R_{\rm eff}$ $\propto$ $M_{\rm bar}^{0.252}$ under the same fit. The scatter decomposes as $\sigma_{\rm obs}$ = 0.211 dex, $\sigma_{\rm meas}$ = 0.094 dex, $\sigma_{\rm int}$ = 0.189 dex; roughly 80 per cent of the observed variance is intrinsic. The median is log $\lambda$ = -7.12.

Two consequences follow. First, this tight structural sequence carries no dynamical information by itself. Second, and more important for what follows: because 91 per cent of the variance in $\lambda$ is shared with mass, any correlation between $\lambda$ and RAR residuals is, until proven otherwise, a correlation with mass wearing different units. Every physically meaningful test in this paper is therefore conducted under mass control (partial correlations, mass-quartile permutation nulls, mass-only cross-validation baselines, hierarchical models with mass-correlated random effects). Where $\lambda$ appears without mass control, it is a bookkeeping coordinate, not an inference.

\subsection{Pure global compactness: a boundary test}

The pure global model M1 performs far worse than the RAR baseline ($\Delta$BIC(M1-M0) = +1904.6; Fig. 2). Compactness alone cannot replace the local acceleration structure encoded by the RAR.

\begin{figure}[htbp]
\centering
\includegraphics[width=\linewidth]{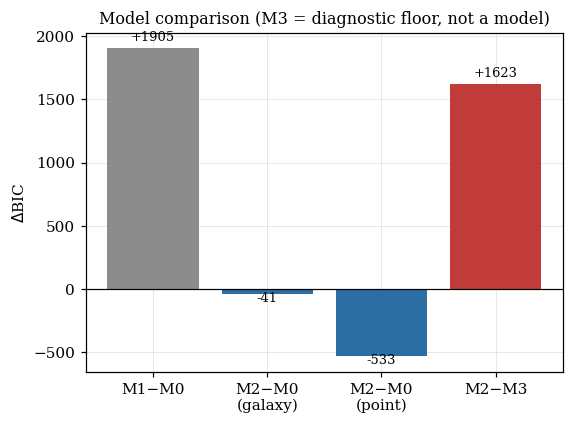}
\caption{Model-comparison $\Delta$BIC ladder (M1--M0, M2--M0 at galaxy and point level, M2--M3); M3 is a diagnostic floor, not a model.}
\label{fig:2}
\end{figure}

\subsection{The hybrid model improves the fit but not the prediction}

M2 improves on the RAR baseline at both levels: $\Delta$BIC(M2-M0) = -41.1 at the galaxy level and -532.6 at the point level, with best-fit $\varepsilon$ = -0.077, $\gamma$ = -0.629. Read jointly, the signs mean the modulation suppresses predicted accelerations most strongly in the most diffuse galaxies: with the best-fit parameters, the factor [1 + $\varepsilon$($(\lambda/\lambda_0)^{\gamma}$] evaluates to $\approx$0.73 at the 10th percentile of compactness and $\approx$0.98 at the 90th. Diffuse galaxies sit below the RAR while compact galaxies track it (mean galaxy residuals of -0.29 and -0.02 dex in the extreme $\lambda$ deciles), which is the same pattern the positive $\lambda$--residual correlation expresses. That asymmetry is consistent with pressure support as a confound: the most diffuse gas-rich dwarfs carry the largest unmodelled pressure-support terms, which lower observed velocities and produce exactly the negative residuals the $\lambda$ term absorbs. One property of the multiplicative form should be stated: with $\varepsilon$ < 0 the factor [1 + $\varepsilon$($(\lambda/\lambda_0)^{\gamma}$] is positive only above a critical compactness, log $\lambda$ = -8.77 for the frozen fit. All 126 galaxies lie above it, the least compact (CamB, log $\lambda$ = -8.67) by 0.10 dex, so the model is well defined throughout the sample --- though the factor reaches 0.13 for that galaxy, an aggressive suppression that an additive log-space form would avoid --- and the form should not be extrapolated to more diffuse systems than SPARC contains. Because $\varepsilon$ and $\gamma$ are strongly degenerate near $\lambda_0$ (the modulation depends on their product over the sampled $\lambda$ range), the fitted values should be read jointly, not separately; a two-dimensional confidence region is provided in the supplementary material. A direct consequence of this degeneracy is that different objective functions select different points along a common ridge rather than a unique minimum; two fits can therefore disagree on ($\varepsilon$, $\gamma$) individually while describing the same modulation. Supplementary Figure S1 shows this concretely: under the per-point error model declared in Section 2.6 the $\chi^{2}$ minimum sits at ($\varepsilon$, $\gamma$) = (-0.042, -0.77), while the frozen-pipeline objective returns (-0.077, -0.63); both lie on the same ridge, and both sit inside the 95 per cent region of a galaxy-level bootstrap (2,000 resamples of whole galaxies, which respects the within-galaxy correlation of distance and inclination errors). We quote uncertainties from that bootstrap: 68 per cent intervals $\varepsilon$ $\in$ [-0.066, -0.023] and $\gamma$ $\in$ [-0.95, -0.64]. Point-level $\Delta\chi^{2}$ contours are several times tighter and are underestimates for exactly the correlated-error reason given in Section 2.5; we do not use them for inference. A non-negative $\varepsilon$ occurred in 1 of 2,000 bootstrap resamples, so the sign of the modulation is robust; whether that modulation is structural physics or absorbable nuisance is the subject of the remainder of this paper, and the galaxy-compressed comparison remains the primary inference.

\begin{figure}[htbp]
\centering
\includegraphics[width=\linewidth]{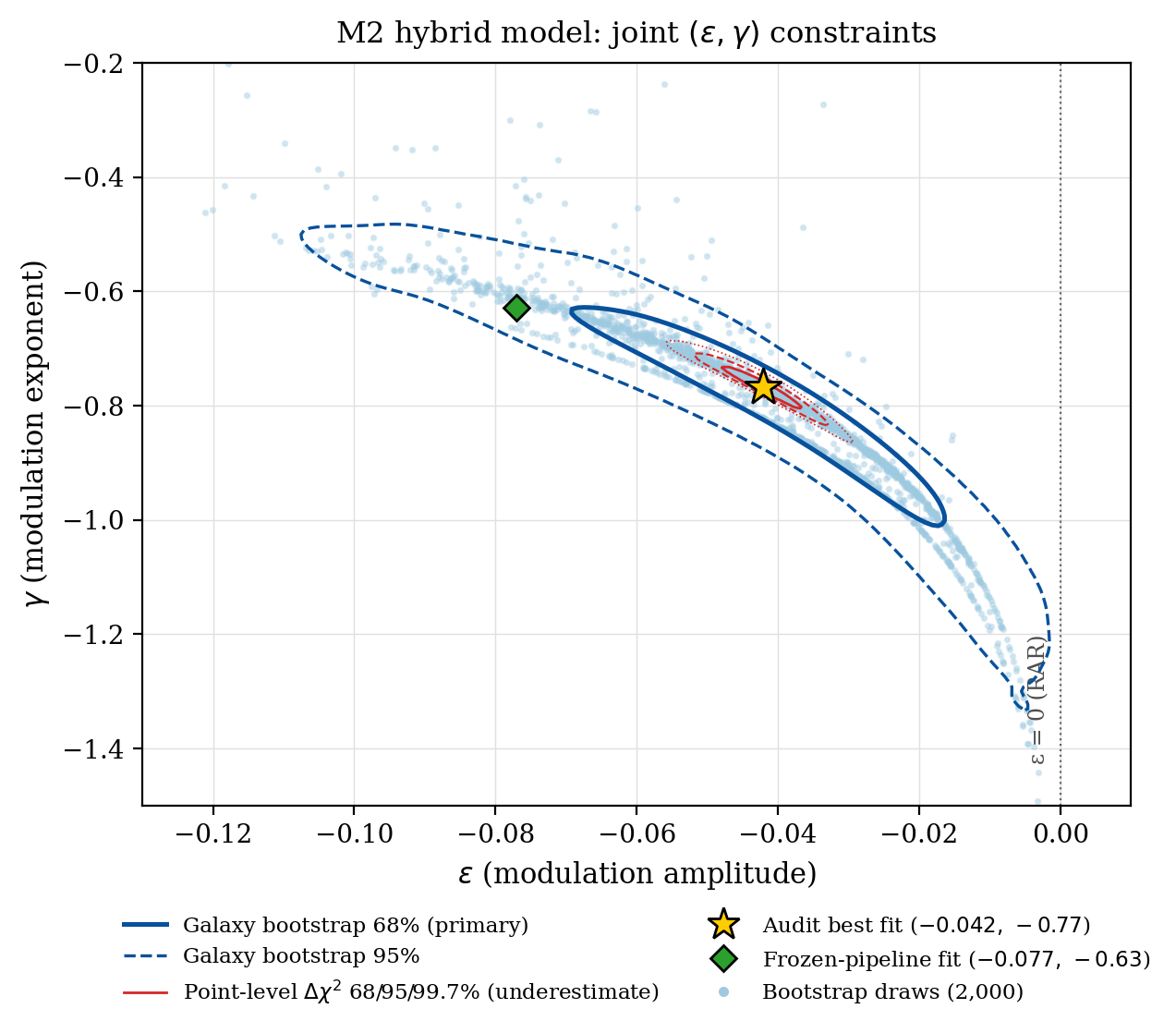}
\caption{Joint ($\varepsilon$, $\gamma$) constraints for the M2 hybrid model, $g_{\rm obs}$ = $g_{\rm RAR}$($g_{\rm bar}$) $\times$ [1 + $\varepsilon$($(\lambda/\lambda_0)^{\gamma}$], $\lambda_0$ = 10$^-$. Blue: galaxy-level bootstrap draws (2,000 resamples of whole galaxies) with 68 and 95 per cent highest-density regions --- the primary uncertainty statement, since it respects within-galaxy correlated distance and inclination errors. Red: point-level $\Delta\chi^{2}$ joint contours (2.30 / 6.17 / 11.83 for two parameters), shown to quantify how severely correlated errors are underestimated at point level. The audit best fit under the declared per-point error model (star, -0.042, -0.77) and the frozen-pipeline fit (diamond, -0.077, -0.63) lie on the same degeneracy ridge, $\varepsilon$($(\lambda/\lambda_0)^{\gamma}$ $\approx$ const over the sampled compactness range. A non-negative $\varepsilon$ occurs in 1 of 2,000 bootstrap draws. The galaxy-compressed likelihood comparison ($\Delta$BIC = -41.1) remains the primary inference.}
\label{fig:S1}
\end{figure}

The nuisance stress test M3 reduces residual scatter from the M0 baseline of $\sigma_{M0}$ = 0.186 dex to $\sigma_{M3}$ = 0.106 dex, consistent with the per-galaxy budget expected from known distance, inclination, and mass-to-light uncertainties; we adopt $\sigma_{M3}$ = 0.106 dex as the nuisance floor throughout, and its agreement with the independently predicted ~0.10 dex budget of Section 1 is itself a corroboration. The scatter reduction from M0 to M2 is 0.015 dex; from M0 to M3 it is 0.080 dex of absorbable per-galaxy scatter. In variance terms, the structural term accounts for 15.5 per cent of the M0 residual variance while per-galaxy zero-point freedom accounts for 67.6 per cent --- the structural gain is 23 per cent of the variance that nuisance freedom absorbs --- and $\Delta$BIC(M2-M3) = +1623 at the point level. As stated in Section 2.5, that ordering is guaranteed by construction; the numbers that matter are the floor and the absorbable scatter, which any one-parameter gain must be judged against.

\textbf{Cross-validation against three baselines.} BIC comparisons reward in-sample fit. The sharper question is predictive: does knowing a galaxy's compactness help predict its RAR residual out of sample, and relative to what? We perform leave-one-out cross-validation at the galaxy level, scoring mean squared prediction error (MSPE) on the held-out galaxy's mean fixed-protocol residual, against three baselines: (i) the RAR itself, predicting zero residual for every galaxy; (ii) a mass-only model, fit as $\delta$ = a + b log $M_{\rm bar}$ on the training set; (iii) a quality-flag baseline, predicting the training-set mean residual of galaxies sharing the held-out galaxy's Q flag.

\begin{figure}[htbp]
\centering
\includegraphics[width=\linewidth]{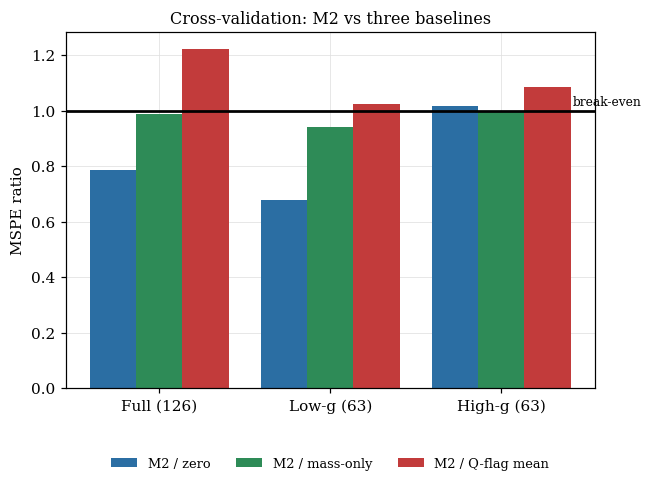}
\caption{Leave-one-out cross-validation MSPE ratios for the hybrid compactness model (M2) against three baselines: the zero-residual RAR prediction, a mass-only model, and a quality-flag (Q-mean) baseline, in the full, low-acceleration, and high-acceleration samples. A bold line marks the ratio = 1.0 break-even threshold. Values below 1.0 indicate superior out-of-sample prediction by M2. In the full sample M2 fails to outperform the mass-only baseline (0.986) and is outperformed by the Q-mean baseline (1.220): apparent structural predictive power is carried almost entirely by baryonic mass and data quality. The figure demonstrates predictive non-identifiability; it does not demonstrate that the RAR is physically exact.}
\label{fig:3}
\end{figure}

Against the zero baseline (Fig.~\ref{fig:3}), M2 reduces prediction error by 21.4 per cent in the full sample (ratio 0.786) and 32.2 per cent in the low-acceleration subsample (0.678), with no improvement in the high-acceleration subsample (1.017). But the zero baseline is the weakest of the three. Against the mass-only baseline, M2's advantage nearly vanishes: MSPE ratio 0.986 in the full sample and 0.942 in the low-acceleration subsample. Against the quality-flag baseline, M2 loses in the full sample (ratio 1.220) and is marginal in the low-acceleration subsample (1.024). Two conclusions follow. First, most of the apparent predictive power of compactness is mass. Second, the SPARC quality flag alone is a better full-sample predictor of RAR residuals than the structural model, which quantifies, more directly than any correlation table, how deeply data quality is entangled with the residual structure of the sample. These results define the extraction limit empirically: a structural correction that cannot out-predict a mass-only baseline, and is beaten by a rudimentary data-quality proxy, is not identifiable as structural in this sample.

Partial correlations tell the same story. Controlling for log $M_{\rm bar}$, the $\lambda$--residual correlation is not significant in the full sample (r = 0.110, p = 0.22) and is marginally significant in the low-acceleration subsample (r = 0.287, p = 0.023).

\subsection{Sensitivity to the mass-modelling strategy}

Under the tuned protocol the full-sample structure--residual correlation is strong ($r_{\rm tuned}$ = 0.700, slope = 0.173 $\pm$ 0.016). Under the fixed protocol it weakens to $r_{\rm fixed}$ = 0.326 (slope = 0.086 $\pm$ 0.022, p = 2.0 $\times 10^{-4}$): a 53.5 per cent reduction. The attenuation combines two effects, removal of tuning covariance (which inflates the tuned correlation) and classical regression dilution (fixing $\Upsilon_{*}$ converts intrinsic stellar-population scatter into unmodelled mass error, biasing the correlation toward zero). The fixed-protocol correlation is therefore a conservative lower bound on any genuine association. Within the symmetric radius policy the correlation is stable against the radius operator ($R_{\rm disk}$: r = 0.323).

\subsection{Regime localization}

\begin{figure}[htbp]
\centering
\includegraphics[width=\linewidth]{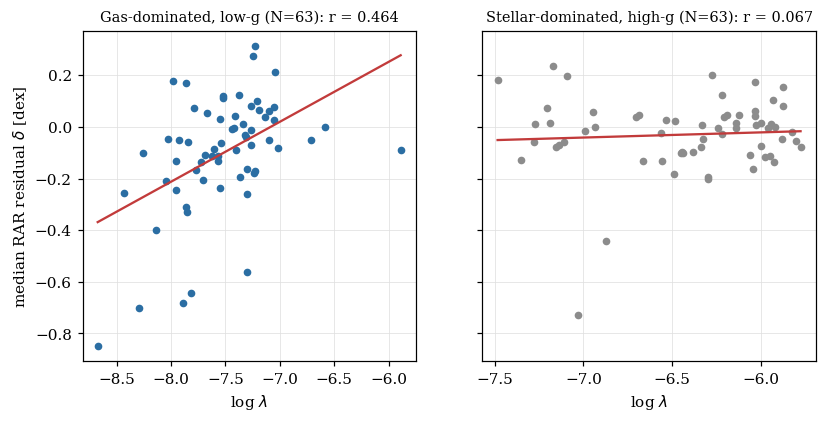}
\caption{Regime localization: fixed-protocol residual vs compactness in the gas-dominated and stellar-dominated halves.}
\label{fig:4}
\end{figure}

The gas-dominated, low-acceleration regime is where the phenomenological question sharpens. Split at the median log $g_{\rm bar}$ = -10.926, the structural residual separates cleanly along the acceleration axis (Fig.~\ref{fig:4}): significant in the gas-dominated half, null in the stellar-dominated half.

\textbf{Gas-dominated, low-acceleration systems} (N = 63): $r_{\rm fixed}$ = 0.464, p = 1.3 $\times 10^{-4}$; $r_{\rm tuned}$ = 0.549; $\Delta$BIC(M2-M0) = -412.4.

\textbf{Stellar-dominated, high-acceleration systems} (N = 63): $r_{\rm fixed}$ = 0.067, p = 0.60; $r_{\rm tuned}$ = 0.448; $\Delta$BIC(M2-M0) = +4.6.

Under the fixed protocol the structural residual is significant in gas-dominated systems and null in stellar-dominated systems. One caveat belongs here rather than in a footnote: with N = 126 the full-sample partial correlation (r = 0.11, p = 0.22) is not statistically distinguishable from the low-acceleration value (r = 0.29) at these sample sizes, so the regime split localizes the signal without proving that the effect is absent elsewhere; a small sample-wide effect remains compatible with the data.

\begin{figure}[htbp]
\centering
\includegraphics[width=\linewidth]{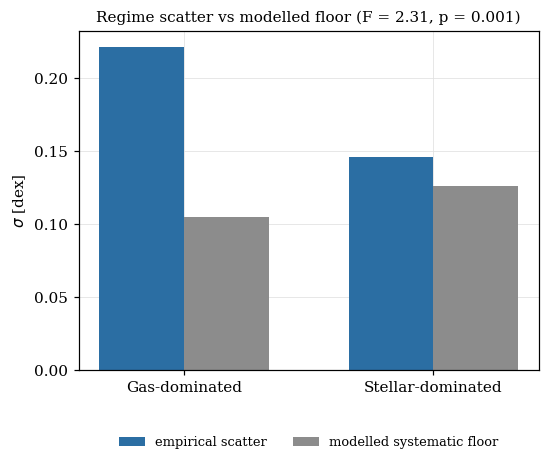}
\caption{Regime scatter against the modelled systematic floor.}
\label{fig:5}
\end{figure}

\textbf{Variance decomposition.} In the stellar-dominated bin (Fig.~\ref{fig:5}) the empirical per-galaxy scatter is $\sigma$ = 0.146 dex, consistent with the modelled systematic floor (~0.126 dex from $\Upsilon_{*}$, distance, and inclination; ratio 1.16). In the gas-dominated bin the empirical scatter is $\sigma$ = 0.221 dex, 2.1 times the expected floor (~0.105 dex); a variance-ratio F-test gives F = 2.31, p = 0.001. The stellar-dominated regime is fully explained by known systematics, which is why no signal survives there. The gas-dominated regime carries excess variance that known systematics do not explain, and the structural residual lives inside that excess. The natural rebuttal --- that low-surface-brightness dwarfs simply carry intrinsically larger error budgets, inflating the apparent excess --- fails a direct test: propagating each galaxy's own distance and inclination uncertainties through the local RAR slope yields per-galaxy floors that are statistically identical in the two regimes (median 0.070 dex in both), the per-galaxy floor does not correlate with residual amplitude (r = -0.08, p = 0.54), and controlling for it leaves the localized correlation unchanged (r = 0.45, p = 2 $\times 10^{-4}$). The regime asymmetry is a property of the residuals, not of the error budgets. Whether the excess is physics or unmodelled kinematics is the question the rest of the paper fails to settle, and says so.

This decomposition also locates our result relative to the SPARC team's own analyses. \citet{Lelli2017} and \citet{Li2018} report that RAR scatter is consistent with observational error alone, leaving no room for a structural residual. Our high-acceleration result agrees with theirs: the modelled floor reproduces the empirical scatter. Our gas-dominated result identifies excess variance beyond the systematics modelled in those papers. The disagreement, such as it is, is confined to the regime where their error model and ours are both least certain.

\begin{figure}[htbp]
\centering
\includegraphics[width=\linewidth]{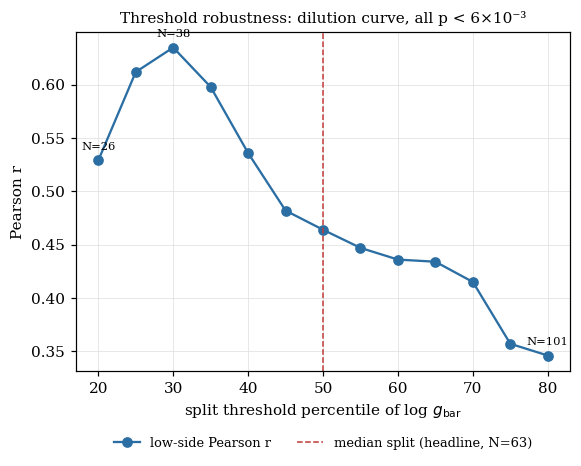}
\caption{Threshold robustness: the low-side correlation across the 20th--80th percentile split range, with sample sizes.}
\label{fig:6}
\end{figure}

\textbf{Threshold robustness.} The median split is a data-driven partition, not a physical boundary. Sweeping the split threshold across the full 20th to 80th percentile range (Fig.~\ref{fig:6}) of log $g_{\rm bar}$, the low-acceleration subsample retains a significant correlation at every threshold (r = 0.35--0.64, all p < 6 $\times 10^{-3}$; peak r = 0.635 at the 30th percentile, N = 38; r = 0.346 at the 80th, N = 101), while the high-acceleration complement remains insignificant. The signal weakens monotonically as the subsample dilutes toward the full sample, exactly as a localized effect should. Because the correlation is a smooth function of the cut, independently implemented variants of this analysis with shallower thresholds or slightly different sample construction (e.g. N = 83 rather than 63) recover correspondingly diluted values (r $\approx$ 0.38) on the same curve; alternate values of this kind, including one archived with the reproducibility record, are cut-placement effects, not inconsistencies. Spearman rank correlation at the median split ($\rho$ = 0.457, p = 1.6 $\times 10^{-4}$) confirms the result is not parametric, and bootstrap resampling (10,000 iterations) gives a 95 per cent CI of [0.22, 0.69] on the low-g Pearson r. The result is also insensitive to the treatment of the RAR zero point itself: refitting the characteristic acceleration to the fixed-protocol sample (g$\dagger$ = 9.3 $\times 10^{-11}$ m s$^{-2}$, rather than adopting the literature $a_0$ = 1.2 $\times 10^{-10}$) leaves the low-acceleration correlation essentially unchanged (r = 0.462 vs 0.464), so the localized residual is not an artefact of the adopted RAR normalisation. Nor is it an artefact of the fixed stellar-mass anchor: substituting independent WISE W1 colour-based stellar masses (\citealp{Duey2025}; 85 sample galaxies in common, median offset +0.10 dex, MAD scatter 0.18 dex against the fixed-$\Upsilon$ masses) into the compactness coordinate leaves the correlation unchanged (r = 0.459 vs 0.464). Because the substitution enters only the structural coordinate --- the rotation-curve decompositions retain the fixed-$\Upsilon$ masses --- this test validates the stability of the structural ranking on the mass--size manifold; it is not a full re-anchoring of the dynamical models. The coverage of that independent sample carries its own information, discussed in Section 4.4.

\textbf{Gas-fraction split.} Splitting instead on gas fraction at the median log $f_{\rm gas}$ = -0.34 yields r = 0.41 (p = 9.6 $\times 10^{-4}$) in the high-gas subsample, with 85 per cent overlap (58/63) between the low-acceleration and high-gas selections, and a monotonic trend across gas-fraction tertiles (r = 0.35, 0.32, 0.43). The localization tracks gas dominance, not the particular partition statistic. A full-sample interaction model ($\delta$ ~ log $\lambda$ + log $\lambda$ $\times$ log $g_{\rm bar}$) gives a non-significant interaction (p = 0.11), so the split is a localization heuristic, to be tested prospectively on independent samples, not a measured phase boundary.

\subsection{Permutation significance test}

Galaxy-level permutation tests (10,000 iterations) assess whether the localization exceeds structured chance. Under an unconstrained shuffle, the observed low-acceleration correlation is exceeded in fewer than 1 in 10,000 iterations (p < $10^{-4}$). Under a mass-controlled null that shuffles residuals within baryonic-mass quartiles, the low-acceleration correlation remains significant (p < $10^{-3}$) and the low-versus-high contrast ($\Delta$r = 0.39) is exceeded in fewer than 1 per cent of iterations (p = 0.009). The localization is not an artefact of sample structure, mass--compactness covariance, or the split procedure, within the nulls tested.

\subsection{Quality-controlled sub-sample removal and the inclination cut}

The low-acceleration, gas-dominated domain is among the most kinematically challenging in SPARC. We therefore repeat the low-acceleration analysis after removing the 10 galaxies most vulnerable to each of six kinematic-quality proxies in turn. Removing the most face-on galaxies strengthens the correlation (r = 0.46 $\rightarrow$ 0.53), as does removing the highest inclination-uncertainty systems (0.53). Removing the smallest angular-extent systems (0.46), most extreme gas fractions (0.47), fewest rotation-curve points (0.46), or lowest quality flags (0.47) leaves it unchanged. Removing the eight galaxies flagged by three or more proxies simultaneously gives r = 0.47 (p = 2.5 $\times 10^{-4}$). The only sub-sample removal that materially weakens the signal is removing the 10 largest-|$\delta$| galaxies (r $\rightarrow$ 0.22), which tests signal concentration, not quality: high-residual galaxies dominate any residual correlation by construction.

\textbf{Inclination cut.} Because most SPARC analyses adopt i > 40$^{\circ}$, we re-run the headline analysis at that cut. The sample drops to N = 113; the low-acceleration correlation strengthens to r = 0.522 (p = 3.1 $\times 10^{-5}$, N = 57), the high-acceleration null persists (r = 0.06, p = 0.66), and the canonical joint partial correlation of Section 4.4 survives (r = 0.328, p = 0.017). The wider 30$^{\circ}$ cut adopted for the frozen sample is therefore conservative: the marginal 30--40$^{\circ}$ galaxies dilute rather than drive the signal.

Robust statistics confirm the signal is broad. Theil--Sen regression gives a slope of 0.216 (95 per cent CI [0.101, 0.340]), close to the OLS slope of 0.232 and excluding zero. Winsorized correlations (5 and 10 per cent) give r = 0.48 and 0.47; a biweight midcorrelation gives 0.42. Aggressive trimming weakens but does not kill the correlation (removing the five largest-|$\delta$| galaxies: r = 0.33, p = 0.011). Controlling individually for the quality flag ($r_{\rm partial}$ = 0.342, p = 0.006), inclination (0.446), or number of points (0.436) leaves it intact. Sequential removal by residual magnitude shows a smooth decline (top 5 removed: $\rho$ = 0.39, p = 0.002; top 10: $\rho$ = 0.28, p = 0.04), so the signal is partly concentrated in high-amplitude systems, as any residual correlation must be, but does not vanish abruptly.

\subsection{Leave-k-out stability}

Randomly removing k galaxies from the low-acceleration subsample (2,000 draws per depth): at k = 5, 99.8 per cent of draws remain significant at p < 0.01; at k = 10, 96.7 per cent; at k = 15 (a 24 per cent reduction), 90 per cent. No single galaxy contributes more than $\Delta$r = 0.09. The signal is broad, not fragile.

\subsection{Residual scatter profile}

\begin{figure}[htbp]
\centering
\includegraphics[width=\linewidth]{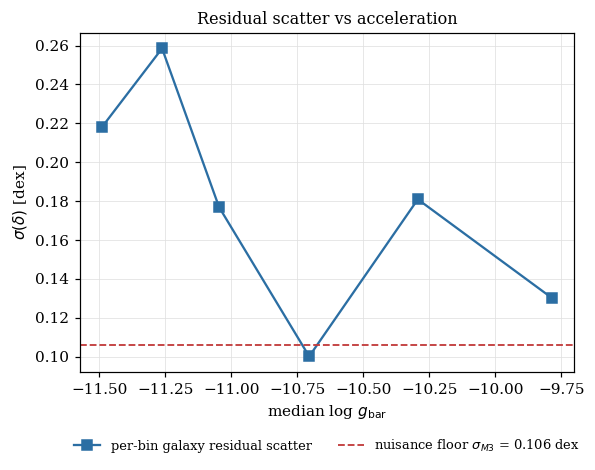}
\caption{Galaxy-level residual scatter vs acceleration, with the $\sigma_{M3}$ = 0.106 dex nuisance floor.}
\label{fig:7}
\end{figure}

Galaxy-level residual scatter varies strongly with acceleration (Fig.~\ref{fig:7}). In the four lowest-acceleration bins the per-galaxy scatter is 0.20--0.26 dex, well above the M3 floor of 0.106 dex; in the two highest bins it drops to 0.07--0.15 dex, approaching the floor. The regime-specific BIC comparison mirrors this: M2 improves the fit where excess variance exists ($\Delta$BIC = -412.4, low) and not where it does not (+4.6, high).

\subsection{Pressure support: the bracket}

Pressure support in gas-rich dwarfs is the dominant untested kinematic systematic, and we bracket it from both sides rather than reporting a single favorable case.

\begin{figure}[htbp]
\centering
\includegraphics[width=\linewidth]{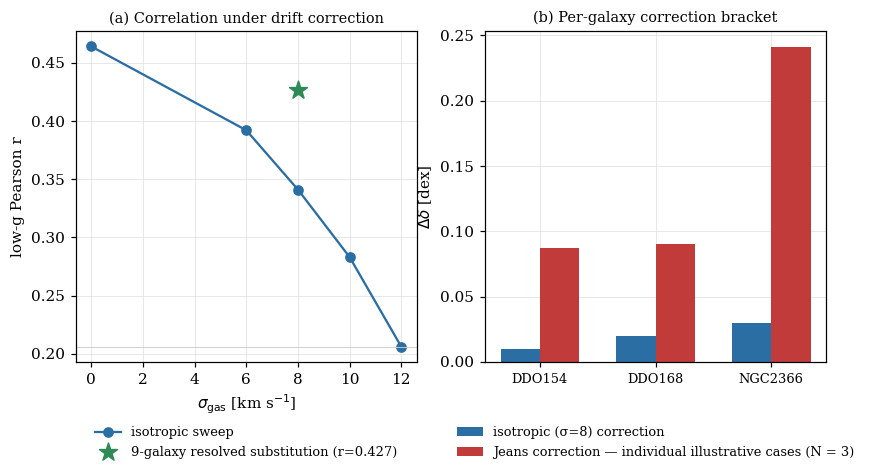}
\caption{Asymmetric-drift correction bounds for the low-acceleration regime. The isotropic flat-dispersion sweep ($\sigma_{\rm gas}$ = 6--12 km/s) shows the progressive absorption of the compactness--residual correlation under the lower-bound treatment. Overplotted with distinct markers (stars, not circles) are the full Jeans-equation corrections for the only three galaxies with resolved LITTLE THINGS dispersion profiles (DDO154, DDO168, NGC2366); the legend labels these explicitly as individual illustrative cases (N = 3), not a fitted cohort --- they are a methodological warning, not a sample sweep, and the Jeans treatment overcorrects two of the three. The gap between the two treatments exceeds the residual amplitude, which is the figure's central point.}
\label{fig:8}
\end{figure}

The lower bound (Fig.~\ref{fig:8}) is the isotropic, flat-dispersion limit, $V_{\rm circ}^{2}$ = $V_{\rm rot}^{2}$ + 2$\sigma_{\rm gas}^{2}$. Applied at the literature-standard $\sigma_{\rm gas}$ = 8 km/s (\citealp{Oh2015}; \citealp{Iorio2017}), the low-acceleration correlation weakens from r = 0.464 to r = 0.341 (p = 0.006; Spearman $\rho$ = 0.363), absorbing roughly 27 per cent of the signal amplitude; residual scatter drops from 0.226 to 0.177 dex, approaching the M3 floor. The correction scales as expected with dispersion: r = 0.392 at $\sigma$ = 6, r = 0.283 at $\sigma$ = 10, and non-significance at $\sigma$ = 12 km/s (r = 0.206, p = 0.10). The largest shifts (>0.1 dex) occur in the slowest rotators ($V_{\rm rot}$ < 25 km/s).

An illustrative upper extreme --- not a statistical population bound --- is the full Jeans equation, $V_{\rm circ}^{2}$ = $V_{\rm rot}^{2}$ - R$\sigma^{2}$ d/dR[ln($\Sigma_{\rm gas}$ $\sigma^{2}$)], evaluated on the three SPARC galaxies with resolved LITTLE THINGS moment maps (\citealp{Hunter2012}): DDO154, DDO168, and NGC2366. There the corrections are far larger, +0.087, +0.090, and +0.241 dex respectively, because the steeply declining $\Sigma_{\rm gas}$ profile contributes substantially. This treatment overcorrects two of the three galaxies, pushing DDO154 and NGC2366 to positive residuals, which indicates that its thin-disc and isotropy assumptions overestimate the correction in dwarf irregulars.

With three galaxies, the Jeans result is a methodological warning about the variance of thin-disc, isotropic assumptions in dwarf irregulars, not a statistical bound. A partial resolved cross-check is already possible: a 1-arcminute positional match to LITTLE THINGS recovers nine overlap galaxies rather than three (six are hidden behind naming differences, e.g. DDO 50 = UGC04305, WLM = UGCA444), all in the low-acceleration subsample, with published asymmetric-drift-corrected rotation curves (\citealp{Oh2015}). Substituting the residuals implied by those curves for the nine galaxies reduces the low-acceleration correlation only mildly (r = 0.464 $\rightarrow$ 0.427, p = 4.9 $\times 10^{-4}$), between the raw and isotropic $\sigma$ = 8 values; a leverage decomposition shows the shift is carried almost entirely by a single galaxy (D564-8, whose +0.29 dex correction alone accounts for $\Delta$r = -0.042), so the substitution constrains but cannot by itself characterize the pressure-support contribution. This substitution is a consistency check, not a measurement: two of the nine shifts are negative, which pressure support cannot produce, so the differences conflate the drift correction with independent-pipeline effects (tilted-ring fitting, inclination, distance scale). It does indicate that the three-galaxy Jeans treatment overestimates: for NGC2366 the published resolved correction is +0.06 dex against the +0.24 dex thin-disc Jeans value. The true correction lies between the two treatments, and the honest summary is uncomfortable in both directions: the methodological uncertainty in the asymmetric-drift correction itself, a factor of several between the isotropic and Jeans limits, exceeds the amplitude of the residual being tested.

\begin{figure}[htbp]
\centering
\includegraphics[width=\linewidth]{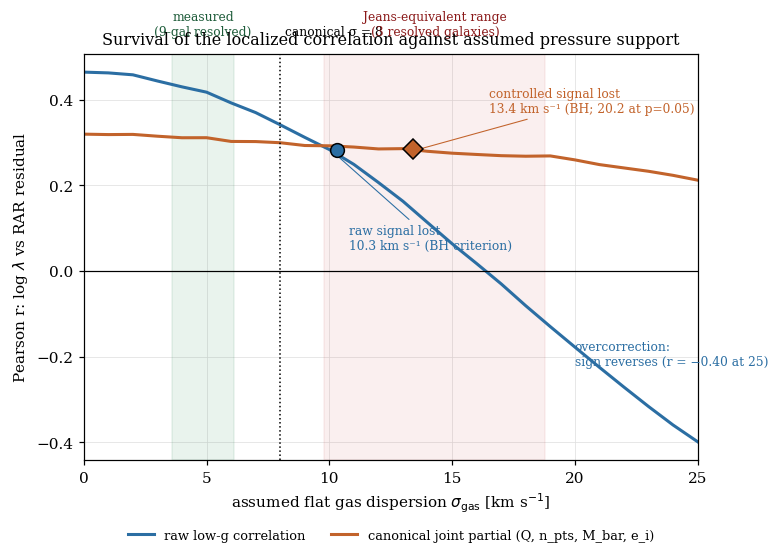}
\caption{Survival of the compactness--residual correlation against an assumed isotropic gas dispersion. The raw low-acceleration correlation (blue) falls below the uniform significance criterion (the canonical test's Benjamini--Hochberg threshold, p = 0.0296) at $\sigma_{\rm gas}$ = 10.3 km s$^{-1}$; the canonical joint partial (orange) at 13.4 km s$^{-1}$ (11.0 and 20.2 under a nominal p = 0.05 criterion). Shaded bands mark the measured nine-galaxy resolved equivalent (3.6--6.1 km s$^{-1}$, green) and the three-galaxy Jeans-equivalent range (9.8--18.8 km s$^{-1}$, red). Past $\sigma_{\rm gas}$ $\approx$ 16.5 km s$^{-1}$ the raw correlation reverses sign, demonstrating that the flat-dispersion device is a bracket, not an estimator.}
\label{fig:9}
\end{figure}

The bracket endpoints do not by themselves say where the signal is lost, so we state it: sweeping the flat dispersion continuously from 0 to 25 km s$^{-1}$ (Fig.~\ref{fig:9}), the raw low-acceleration correlation falls below the canonical test's Benjamini--Hochberg threshold (p = 0.0296, adopted as a uniform criterion) at $\sigma_{\rm gas}$ = 10.3 km s$^{-1}$, and the canonical joint partial at 13.4 km s$^{-1}$ (under a nominal p = 0.05 criterion these become 11.0 and 20.2 km s$^{-1}$ --- the boundary is criterion-dependent and we quote the conservative values). For scale, the resolved nine-galaxy corrections of Section 3.10 correspond to equivalent flat dispersions of only 3.6 (median) to 6.1 (mean) km s$^{-1}$, while the three-galaxy Jeans values map to 9.8--18.8 km s$^{-1}$. The signal therefore survives the corrections that have been measured and dies inside the theoretical range that cannot be excluded --- the paper's identifiability verdict restated in physical units. Pushing the flat correction further makes the point from the other side: past $\sigma_{\rm gas}$ $\approx$ 16.5 km s$^{-1}$ the raw correlation changes sign, reaching r = -0.398 at 25 km s$^{-1}$, so overcorrection does not merely erase the signal, it manufactures an artificial one of opposite sign. The flat-dispersion device is a boundary bracket, not an estimator, and no single "true" value of $\sigma_{\rm gas}$ should be selected from it. Pressure support therefore cannot be quantified at the required precision with current data, which is precisely why the interpretation of the localized residual remains open. Resolved $\sigma_{\rm gas}$(R) measurements for a statistically useful overlap sample (the THINGS survey offers ~15--20 candidates) are the decisive missing observation.

\subsection{One-parameter structural comparisons}

To test whether the low-acceleration residual is specific to compactness, we compare four one-parameter summaries head to head: $\lambda$, baryonic surface density $\Sigma_{\rm bar}$ = $M_{\rm bar}$/($\pi R_{\rm eff}^{2}$), gas fraction $f_{\rm gas}$, and the size residual at fixed mass. Fixed-protocol correlations with the mean RAR residual in the low-acceleration subsample (N = 63; the mean is used here so that all four coordinates are compared on the same target scored by the cross-validation of Section 3.3 --- the $\lambda$ value differs from the median-based headline by $\Delta$r = 0.013):

\begin{itemize}
\item $\log\lambda$: $r = 0.477$, $p = 7.7\times10^{-5}$ (95\% CI $[0.24,\,0.68]$)
\item $\log\Sigma_{\rm bar}$: $r = 0.384$, $p = 1.9\times10^{-3}$ (95\% CI $[0.15,\,0.57]$)
\item $\log f_{\rm gas}$: $r = 0.108$, $p = 0.40$
\item $\log R_{\rm resid}$: $r = -0.126$, $p = 0.33$
\end{itemize}

After mass control, the three mass-dependent variables converge to the same partial correlation, |$r_{\rm partial}$| = 0.301 (p = 0.016), and a Steiger test finds the $\lambda$ and $\Sigma_{\rm bar}$ correlations indistinguishable (Z = 1.58, p = 0.12). The partial-correlation framework cannot rank structurally coupled variables; it establishes only that the residual information is shared across the mass--size manifold rather than owned by any coordinate. In the high-acceleration subsample all four variables are null (|r| < 0.12, p > 0.35).

\subsection{Multivariate structural-manifold test}

A multivariate OLS regression of the low-acceleration residual on all four standardised structural variables gives adjusted R$^2$ = 0.224 (F = 6.13, p < 0.001), but with severe collinearity ($\Sigma_{\rm bar}$ and $R_{\rm resid}$ at r = -0.95). Ridge regression at the cross-validated penalty retains log $\lambda$ most strongly; LASSO selects log $\lambda$ and log $f_{\rm gas}$ while zeroing the others ($R^{2} = 0.246$). Neither selection is physically meaningful: the coefficient path shows all four variables entering in rapid succession as the penalty relaxes, so the sparse selection is representation-dependent within a collinear feature set. Principal component analysis makes the geometry explicit: the residual correlates with PC1, the mass--size axis (r = 0.370, p = 0.003), and not with PC2, the gas-richness axis (r = -0.148, p = 0.25); all four components together give $R^{2} = 0.251$, identical to raw-feature OLS. The manifold collectively carries a modest, real signal (22--25 per cent of low-acceleration residual variance) that no single coordinate owns. The LASSO coefficient path is relegated to the supplementary material, since its content is the non-uniqueness already established here.

\section{Discussion}

\subsection{The extraction limit}

The audit's primary result is quantitative and negative: in the full SPARC sample under a fixed protocol, no one-parameter structural correction to the RAR is identifiable above per-galaxy nuisance freedom. The evidential chain is: (i) the per-galaxy floor absorbs 0.080 dex of variance against M2's 0.015 dex gain; (ii) under cross-validation the compactness model cannot out-predict a mass-only baseline (MSPE ratio 0.99) and loses to a quality-flag baseline (1.22); (iii) after mass control the full-sample correlation is null (r = 0.11, p = 0.22). Any of these alone would be suggestive; together they close the question for this sample and protocol. Nor is the closure an artefact of sample size: at N = 126 the analytic power to detect a correlation of r = 0.30 at $\alpha$ = 0.05 is 93 per cent, and 81 per cent for r = 0.25, so the full-sample null is a genuine non-detection rather than an underpowered miss; the localized subsample (N = 63) has 97 per cent power at its observed amplitude. The one regime where power genuinely limits inference is the Q = 1 low-acceleration subsample, as Section 4.4 and Appendix C state. Small hybrid gains in RAR-adjacent work are often discussed as physically suggestive without being benchmarked against per-galaxy freedom or mass-only prediction. The benchmark is cheap, and we suggest it become standard.

Recent analyses report lower intrinsic RAR scatter than the 0.10 dex floor here: \citet{Desmond2023} finds $\sigma_{\rm int}$ = 0.034 dex with hierarchical HMC inference; \citet{Varasteanu2025} measure 0.045 dex in MIGHTEE-HI; \citet{Mistele2024} extend the relation to the weak-lensing regime; \citet{Chae2020} report external-field signatures in SPARC residuals. The extraction boundary measured here is protocol- and sample-specific, and will tighten as data and inference improve. That is the point of measuring it explicitly.

We emphasize that the inability to isolate a unique structural correction does not prove the radial acceleration relation exact. It quantifies a strict observational limit: any genuine secondary physical parameter governing galaxy kinematics in SPARC must have an amplitude above the 0.080 dex per-galaxy nuisance floor, and its variance must be sufficiently orthogonal to baryonic mass to survive mass-controlled cross-validation. Until higher-resolution data lower this systematic floor, the mass--size manifold cannot be used to break the degeneracy between modified gravity and baryonic-feedback interpretations.

\subsection{The structural manifold, not compactness}

This paper does not establish compactness as a necessary or even meaningful dynamical correction, and its architecture reflects that. After mass control, compactness and surface density carry statistically indistinguishable information (Section 3.11); regularised regression cannot stably attribute the signal (Section 3.12); the residual aligns with the mass--size axis of the manifold. The editor's objection that a concentration index proportional to a power of mass must correlate with mass is correct, and Section 3.1 quantifies it (91 per cent shared variance). Our claims survive it because none of them rests on raw $\lambda$ correlations: the localized residual is established under mass-quartile permutation nulls, mass-partial correlations, mass-only cross-validation baselines, and hierarchical models. What remains after mass control is a modest, regime-localized, manifold-level signal, and that is all we claim.

\subsection{Gas-dominated systems: cleanest masses, dirtiest kinematics}

The gas-dominated, low-acceleration regime is the one domain where a structural residual survives the extraction limit: r = 0.464 under the fixed protocol, cross-validation ratio 0.678 against zero (though only 0.942 against mass-only), excess variance 2.1 times the modelled floor. It is also the domain where MOND and $\Lambda$CDM feedback models genuinely disagree: a strictly local RAR admits no secondary structural term, while feedback models predict their strongest imprints exactly here. The quality entanglement documented below prevents this paper from discriminating between the frameworks. If the residual is physical, it is consistent with feedback predictions; if it is pressure support, the RAR's strict locality survives and the correlation is an artefact of unmodelled kinematics. The contribution is to have localized the domain where the discriminating test can be run, and to have specified the data it requires: resolved $\sigma_{\rm gas}$(R) in a statistically useful sample of gas-dominated dwarfs.

\subsection{Quality entanglement, stated precisely}

The residual correlates with data-quality indicators, and this section states exactly how much survives joint control, in which metric, since the answer is metric-dependent.

Quality-flag stratification: in the full sample, Q = 1 galaxies (N = 74) give $\rho$ = 0.02 (p = 0.89) while Q = 2 galaxies (N = 44) give $\rho$ = 0.46 (p = 0.002). In the low-acceleration subsample: Q = 1 (N = 31) gives $\rho$ = 0.29 (p = 0.11); Q = 2 (N = 27) gives $\rho$ = 0.54 (p = 0.004). The Q = 1 low-acceleration null removes any claim of a quality-invariant signal. At N = 31 the power to detect r = 0.46 is roughly 75 per cent, and a Fisher z-test cannot distinguish the Q = 1 and full low-acceleration correlations (p = 0.38), so the null is ambiguous between absence and low power. It is nevertheless a boundary condition: no physical interpretation should be advanced that requires the signal to be quality-invariant, because it is not shown to be.

An independent partition reproduces the same boundary. Cross-matching to the WISE-based stellar-mass sample of \citet{Duey2025} splits the low-acceleration subsample into a brighter, predominantly Q = 1 half that WISE covers (N = 28, median log $M_{\rm bar}$ = 9.35) and a fainter, gas-richer, predominantly Q = 2 half that it does not (N = 35, median log $M_{\rm bar}$ = 8.88). The correlation is null in the covered half (r = 0.09, p = 0.64) and carried by the uncovered half (r = 0.46, p = 0.006). The two halves are physically distinct populations, not merely differently sized samples: the uncovered galaxies are fainter (median $\log M_{*}$ = 8.07 vs 8.83; Mann--Whitney p = 3 $\times 10^{-3}$) and, most strongly, lower in effective surface brightness (median log $\Sigma_{\rm eff}$ = 1.25 vs 1.52; p = 1 $\times 10^{-4}$), while their gas fractions do not differ significantly (medians 0.77 vs 0.65, p = 0.21). The partition therefore tracks the infrared survey's surface-brightness selection. The mass anchor itself is validated (Section 3.5); what this partition shows is that the signal resides in the low-surface-brightness dwarf population that independent photometric samples do not reach --- a physical characterization of the host population, though one that coincides with where kinematic extraction is hardest, and with the population flagged by the quality stratification above. The identifiability boundary thus appears from a third, independent angle: every external probe available either cannot separate the signal from a nuisance or cannot reach the galaxies that carry it.

Joint control: correlating log $\lambda$ with the residual while controlling simultaneously for all seven available quality proxies ($e_D$, $e_i$, Q, $n_{\rm pts}$, distance, angular size, inclination) gives, in rank-based (Spearman) partial correlation, r = 0.06 (p = 0.53) in the full sample and r = 0.21 (p = 0.13) in the low-acceleration subsample; in Pearson partial correlation on the same covariates, r = 0.15 (p = 0.11) and r = 0.29 (p = 0.03) respectively. The rank-based control is the more conservative and we quote it first; the divergence between the two metrics is itself informative, indicating that the parametric survival draws support from the distribution tails that rank statistics discount, and the joint-control result should not be summarised by either number alone. The joint control is not collinearity-degraded: the standardized seven-covariate design matrix has condition number 4.4 in the full sample and 5.0 in the low-acceleration subsample, and within the canonical control set the variance inflation factors are 1.2--1.5 for the nuisance covariates, rising to 5.8 (mass) and 6.2 (compactness) for the deliberately entangled pair --- as their 91 per cent shared variance predicts, and below the conventional instability threshold of 10.

Because the multiplicity of possible controls invites selective reporting in both directions, we specified one canonical test in advance of computing it --- recorded in an internal, dated audit roadmap (May 2026) included in the archive, whose date is self-asserted rather than externally registered: the partial correlation between log $\lambda$ and the asymmetric-drift-corrected residual ($\sigma_{\rm gas}$ = 8 km/s), controlling simultaneously for quality flag, number of rotation-curve points, baryonic mass, and inclination error, in the low-acceleration subsample. The result is r = 0.30 (p = 0.021, N = 63, df = 57), and it survives the family-wide Benjamini--Hochberg correction and the i > 40$^{\circ}$ cut (r = 0.33, p = 0.017). The correction margin is finite and worth stating plainly: within the declared family the canonical test clears its threshold by a factor of 1.4 (p = 0.0213 against 0.0296 at rank 16 of 27); it would fail at q = 0.02, or if roughly a dozen further tests joined the family, and at q = 0.05 approximately one of the seventeen surviving tests is expected to be a false discovery. The margin is real but not generous --- consistent with everything else reported about this signal. The test's evidential standing does not rest on the provenance of its specification: it enters the declared 27-test family on equal footing with every other member and passes the same correction, so the conclusion is unchanged whether the reader accepts the advance-specification record or treats the test as one control among many. The corresponding full-sample value is r = 0.19 (p = 0.04). The canonical control returns essentially the same partial correlation as mass control alone (r = 0.299 against the 0.301 of Section 3.11) --- informative rather than coincidental: once baryonic mass is controlled, the remaining covariates carry no additional independent variance against the residual, consistent with the variance inflation factors above, where only the mass--compactness pair is elevated. The localized residual therefore survives simultaneous control for mass, the principal quality covariates, and first-order pressure support, in the metric of the headline analysis. It does not survive rank-based control over the widest proxy set at full-sample level, and the difference between those two statements is the honest width of the uncertainty. One limitation of the canonical test itself must be stated: it incorporates only the first-order isotropic drift correction, and Section 3.10 shows the gap between that approximation and a resolved Jeans treatment can reach ~0.15 dex in individual dwarfs --- so survival under canonical control establishes independence from the proxies of kinematic difficulty, not from pressure-supported kinematics themselves.

Hierarchical models place the same question in a likelihood framework. We fit linear mixed models by maximum likelihood --- point residual $\delta$ as response, mean-centred log $\lambda$ as the fixed effect, a Gaussian random intercept per galaxy --- on the complete matched sample (all 126 galaxies, 2,709 points; an earlier version of this analysis ran on a 78-galaxy subset produced by a name-resolution failure in table matching, and is superseded by the full-sample fits reported here, which the archived pipeline regenerates). log $\lambda$ is a significant fixed effect in the full sample ($\beta$ = 0.097, p = 9.8 $\times 10^{-6}$; $\Delta$BIC = -10.3; $\tau$ = 0.186 $\rightarrow$ 0.173 dex; ICC = 0.75) and in the low-acceleration subsample (63 galaxies, 961 points: $\beta$ = 0.232, p = 1.7 $\times 10^{-5}$; $\Delta$BIC = -9.4; $\tau$ = 0.215 $\rightarrow$ 0.188 dex, a 12 per cent reduction; residual $\sigma$ = 0.131 dex; ICC = 0.73), and is null in the high-acceleration subsample (p = 0.36). The hierarchical model shows that, on average, more compact galaxies sit below the RAR even under principled partial pooling. But the recovered term is small against galaxy-to-galaxy variance, and the result does not single out compactness as the relevant variable; detectability under shrinkage is weaker than identifiability under manifold degeneracy, and only the former is demonstrated.

The evidence ranks as follows. The full-sample non-identifiability is the primary conclusion and is robust to everything we tried. The localized residual is a secondary finding: real at the precision the data allow, surviving the canonical control, but living in the same regime where quality, structure, and kinematic difficulty covary, and therefore not attributable. The manifold non-uniqueness constrains what any future explanation may claim. We cannot resolve the localized residual with these data. A larger Q = 1 sample, or spatially resolved H I kinematics in gas-dominated dwarfs, would settle it.

\subsection{Candidate explanations and their observable signatures}

If the residual is physical, candidate mechanisms include halo response to baryonic concentration (\citealp{DiCintio2014}; \citealp{Katz2017}; \citealp{Read2019}) and feedback-sculpted inner profiles (\citealp{Pontzen2012}), predicting correlation with spin proxies or concentration indicators; gas-fraction-dependent equilibrium departures, predicting the residual weakens under resolved asymmetric-drift correction; and feedback history imprinted on inner halos, predicting dependence on star-formation history at fixed mass. If it is systematic, the candidates are beam smearing, predicting correlation with beam-to-diameter ratio and attenuation at higher resolution; pressure support, predicting correlation with $\sigma_{\rm gas}$/$V_{\rm rot}$; and inclination errors coupling to surface brightness, predicting the residual vanishes under kinematic inclinations.

The internal checks available in SPARC constrain three of these channels. Angular size (beam proxy) does not correlate with the residual ($\rho$ = -0.20, p = 0.12), nor does inclination error ($\rho$ = -0.09, p = 0.47), and controlling for angular size, inclination error, and distance leaves the correlation unchanged (partial $\rho$ = 0.44). The pressure-support channel is bracketed, not excluded, by Section 3.10: the isotropic correction absorbs about a quarter of the signal, the Jeans correction overcorrects resolved test cases, and the truth between them cannot be pinned with current data. A baryonic-tracer estimate of the external field ($g_{\rm ext}$ = $\Sigma GM_j/r_{ij}^{2}$ from ALFALFA-selected tracer masses within 10 Mpc, N = 59 low-acceleration galaxies) yields $g_{\rm ext}$/$a_0$ < $10^{-4}$ throughout with no residual correlation (r = 0.16, p = 0.22), though H I-selected tracers underestimate the true field and a group-catalogue test remains future work. Local-environment controls (nearest-neighbour distance, local density) do not absorb the signal either; the $\lambda$--residual partial correlation is unchanged or strengthens under environment control (N = 114 with \citealp{Durbala2020} tracers).

These channels are not mutually exclusive, and the observed residual likely superposes several. The decisive test requires spatially resolved H I kinematics on an independent sample, where beam smearing is controlled, dispersion is measured per galaxy, and inclination is kinematically constrained.

\subsection{Four generic confounds}

Four hazards isolated here apply to any structure--dynamics residual analysis in resolved rotation-curve samples. Per-galaxy mass-to-light tuning injects covariance between structure and residuals; a fixed protocol reduced our correlation by 53.5 per cent. Radius-operator asymmetry introduces ~0.176 dex of scatter when definitions are mixed across protocols. Galaxy-level zero-point freedom absorbs small one-parameter gains unless explicitly benchmarked. And data-quality indicators covary with structural residuals in ways that mimic physics; here the quality flag alone out-predicts the structural model in full-sample cross-validation. Any structural-residual claim in this class of data should demonstrate robustness to all four.

\section{Conclusions}

This paper asks whether structure--dynamics coupling is uniquely recoverable from rotation-curve observables, and answers with an audit rather than a detection. The audit produces four results, each independently testable on future data.

1. \textbf{No one-parameter structural correction to the RAR is identifiable above the galaxy-level systematic floor in the full SPARC sample under a fixed protocol.} The hybrid compactness model improves the in-sample fit ($\Delta$BIC = -41.1 at the galaxy level) but per-galaxy zero-point freedom absorbs 0.080 dex of residual scatter against its 0.015 dex gain, and under leave-one-out cross-validation it cannot out-predict a mass-only baseline (MSPE ratio 0.986) and loses to a quality-flag baseline (1.220). This extraction limit is the central quantitative result, and the three-baseline predictive benchmark is the audit standard we propose for future structural-residual claims.

2. \textbf{A localized residual survives in the gas-dominated, low-acceleration regime, and survives the controls we can construct, but its physical origin is undetermined.} The correlation is r = 0.46 (p = 1.3 $\times 10^{-4}$), strengthens to r = 0.52 under the stricter i > 40$^{\circ}$ cut, holds across the 20th--80th percentile range of split thresholds, survives mass-controlled permutation nulls (p < $10^{-3}$), remains a significant fixed effect under hierarchical partial pooling on the complete matched sample ($\beta$ = 0.232, p = 1.7 $\times 10^{-5}$; 63 galaxies, 961 points), and survives the canonical joint control specified in advance of computation (quality, sampling, mass, inclination error, first-order pressure support; partial r = 0.30, p = 0.021). It is, however, entangled with data quality: it concentrates in Q = 2 galaxies, the Q = 1 subsample is null at low power, and --- distinctly from the canonical control --- the more conservative rank-based joint control over the widest proxy set leaves it marginal. An isotropic asymmetric-drift correction absorbs about a quarter of it; a full Jeans correction overcorrects resolved test cases; the uncertainty in the correction exceeds the signal.

3. \textbf{The residual belongs to the mass--size manifold, not to compactness.} After mass control, compactness and surface density are statistically indistinguishable (Steiger p = 0.12; both partial r = 0.301). Sparse selection among collinear structural variables is representation-dependent, and PCA locates the signal on the mass--size axis. The manifold collectively explains 22--25 per cent of the low-acceleration residual variance. No claim of a preferred structural coordinate is supported.

4. \textbf{Four generic methodological confounds are quantified} for this class of analysis: mass-to-light tuning covariance, radius-operator asymmetry, unbenchmarked per-galaxy freedom, and quality entanglement, the last being strong enough here that the quality flag out-predicts the structural model out of sample.

The discriminating observations are specified, not speculative. If the localized residual replicates in an independent resolved-H I sample with per-galaxy dispersion measurements, the physical interpretation gains ground and strict RAR locality is challenged. If it vanishes under resolved asymmetric-drift correction or kinematic inclinations, the systematic interpretation wins and the RAR's insensitivity to structure is reinforced at a tighter bound. Either outcome is informative, which is the property that makes the localized residual worth resolving. The required data exist or are imminent: THINGS and LITTLE THINGS provide resolved dispersions for a first statistical test, and MIGHTEE-HI (\citealp{Varasteanu2025}) and WALLABY provide independent samples with orthogonal selection.

The paper's empirical contribution is not to have proved a coupling or ruled one out, but to have localized where the question can actually be posed. In the gas-dominated regime, the correlation is there at r = 0.46, survives the canonical control and every sub-sample removal we could construct --- the single control it does not clear is the most conservative rank-based joint control over the widest proxy set (Section 4.4) --- and lives inside a variance excess ($\sigma$ $\approx$ 0.22 dex against a modelled floor of ~0.10 dex) that known systematics cannot account for on their own. It carries the quality caveats stated in Section 4.4, and no reading of it should require quality invariance. But it is not an artifact of sample construction, mass--compactness collinearity, or per-galaxy nuisance freedom; the permutation, hierarchical, and baseline tests close those doors. Whether the excess is halo response, feedback imprint, unmodelled pressure support, or a combination remains an observational question. The regime is small enough to be observationally tractable and dynamically consequential enough to be worth resolving.

We close with the audit standard itself, since it is the transferable product of this work. A claimed structural correction to the RAR should be required to (i) out-predict a mass-only baseline out of sample, (ii) clear an explicit per-galaxy nuisance benchmark, and (iii) survive data-quality stratification. In SPARC, today, no one-parameter structural correction meets all three. One localized candidate meets the first two in a restricted regime and fails only where the data cannot yet answer.

This work marks a boundary for one-dimensional, single-parameter rotation-curve analysis. The failure of the structural correction to clear the nuisance benchmark and the pressure-support bracket is not a null result in the dismissive sense; it is an observability limit that redefines the requirements for the next generation of dynamical tests. Breaking the degeneracy requires moving from one-dimensional scalar projections to resolved $\sigma_{\rm gas}$/V and angular-momentum profiling --- and the dispersion sweep of Section 3.10 states the requirement quantitatively: the controlled signal survives to an equivalent flat dispersion of 13.4 km s$^{-1}$ under the conservative criterion, while the largest measured resolved correction corresponds to 6.1 km s$^{-1}$, so deciding the question requires per-galaxy dispersion measurements accurate at the few-km s$^{-1}$ level across a statistical sample. The benchmark provided here is intended to ensure that, as those data arrive, genuine physical couplings can be cleanly separated from structural covariance.

\section*{Appendix A: Leave-One-Galaxy-Out Stability}

Recomputing $\Delta$BIC(M2-M0) after removing each galaxy in turn, M2 remains preferred in all 126 iterations (mean $\Delta$BIC = -528.4, range [-564.5, -341.6]). No single galaxy reverses the model ranking.

\section*{Appendix B: Threshold Robustness}

Full sweep from the 20th to the 80th percentile of log $g_{\rm bar}$ (13 thresholds, step 5): the low-side correlation is significant at every threshold, r = 0.529 (N = 26, p = 5.5 $\times 10^{-3}$) at the 20th percentile, peaking at r = 0.635 (N = 38, p = 1.8 $\times 10^{-5}$) at the 30th, declining smoothly to r = 0.346 (N = 101, p = 3.9 $\times 10^{-4}$) at the 80th. The high-side complement is insignificant throughout. Spearman at the median split: $\rho$ = 0.457 (p = 1.6 $\times 10^{-4}$). Bootstrap 95 per cent CI: [0.22, 0.69].

\section*{Appendix C: Quality-Flag Supplemental Check}

Restricting the low-acceleration subsample to Q = 1 (N = 31): $\rho$ = 0.294 (p = 0.11), r = 0.203 (p = 0.27); neither significant. At N = 31 the power to detect r = 0.46 at $\alpha$ = 0.05 is ~75 per cent. A Fisher z-test against the full low-acceleration correlation gives p = 0.38. Relaxing to Q $\leq$ 2 (N = 58) recovers the signal: $\rho$ = 0.453 (p = 3.5 $\times 10^{-4}$), r = 0.510 (p = 4.3 $\times 10^{-5}$). The Q = 1 result neither confirms nor refutes the residual; it is consistent with reduced power, and it caps the strength of any claim. The three-baseline cross-validation restricted to Q = 1 galaxies tells the same two-sided story: in the Q = 1 low-acceleration subsample (N = 31), no model outperforms the zero baseline (M2/zero = 0.994; mass-only/zero = 1.059), consistent with the correlation null, while the ordering between models is preserved (M2/mass-only = 0.939). The quality entanglement is therefore not resolved in either direction by the highest-quality data available: the structural term is neither demonstrated nor reduced to a quality artefact at current Q = 1 sample size.

\section*{Appendix D: Low-Acceleration Subsample Properties}

Table~\ref{tab:D1} lists the 63 low-acceleration galaxies sorted by fixed-protocol median RAR residual, with quality flag, inclination, effective radius, baryonic mass, log compactness, median residual, median baryonic acceleration, and number of valid points. Residual amplitude is concentrated at the top of this ordering: removing the five largest-|$\delta$| galaxies reduces the correlation from r = 0.46 to r = 0.33 (p = 0.011), and removing the ten largest reduces it to r = 0.22. This reflects concentration of residual amplitude rather than failure of the quality sub-sample removals --- any residual correlation is carried disproportionately by its largest residuals. Representative subset below; full table in the online archive.

\begin{table}[htbp]
\centering
\caption{Low-acceleration subsample properties, sorted by fixed-protocol median RAR residual. Representative subset of the 63 low-acceleration galaxies; the full table is in the online archive.}
\label{tab:D1}
\small
\begin{tabular}{lcccccccc}
\hline\hline
Galaxy & $Q$ & $i$ (deg) & $R_{\rm eff}$ (kpc) & $\log(M_{\rm bar}/M_{\odot})$ & $\log\lambda$ & $\delta_{\rm fix}$ & $\log g_{\rm bar}$ & $n_{\rm pts}$ \\
\hline
UGC07399    & 1 & 55 & 1.27 & 9.20 & $-7.23$ & $+0.314$ & $-10.99$ & 10 \\
F568-V1     & 1 & 40 & 4.40 & 9.72 & $-7.25$ & $+0.276$ & $-11.21$ & 15 \\
NGC2915     & 2 & 56 & 0.53 & 9.00 & $-7.05$ & $+0.214$ & $-11.48$ & 30 \\
UGC05764    & 2 & 60 & 1.20 & 8.41 & $-7.99$ & $+0.177$ & $-11.31$ & 10 \\
ESO444-G084 & 2 & 32 & 0.75 & 8.33 & $-7.86$ & $+0.169$ & $-11.16$ & 7 \\
F583-1      & 1 & 63 & 3.74 & 9.52 & $-7.37$ & $+0.122$ & $-11.55$ & 25 \\
UGC07603    & 1 & 78 & 0.85 & 8.73 & $-7.52$ & $+0.112$ & $-10.97$ & 12 \\
UGC08490    & 1 & 50 & 1.14 & 9.17 & $-7.21$ & $+0.100$ & $-11.21$ & 30 \\
UGC04305    & 3 & 40 & 1.23 & 9.11 & $-7.30$ & $-0.562$ & $-11.08$ & 22 \\
F563-V1     & 3 & 60 & 5.01 & 9.20 & $-7.82$ & $-0.644$ & $-11.53$ & 6 \\
PGC51017    & 3 & 66 & 1.28 & 8.54 & $-7.89$ & $-0.682$ & $-11.24$ & 6 \\
UGC07577    & 2 & 63 & 0.77 & 7.91 & $-8.30$ & $-0.701$ & $-11.48$ & 9 \\
CamB        & 2 & 65 & 1.21 & 7.73 & $-8.67$ & $-0.850$ & $-11.30$ & 9 \\
\hline\hline
\end{tabular}
\end{table}

\section*{Appendix E: Methodological Framework and Validation Pipeline}

The analysis follows one governing principle: maximize physical information while minimizing assumptions. Four rules implement it. Start from observations; introduce as few free parameters as possible; validate every transformation statistically; and reject methods that improve appearance rather than predictive power.

The computational workflow, applied end to end in the archived pipeline, is: (1) import and hash-verify the source catalogues; (2) apply quality control and unit-consistency checks; (3) propagate measurement uncertainties, including their covariance structure (distance and inclination errors enter multiple derived quantities simultaneously); (4) construct derived physical quantities and dimensionless variables; (5) fit with error-in-variables methods under a declared error model; (6) decompose observed scatter into measurement and intrinsic components; (7) validate with Monte Carlo perturbation, bootstrap resampling, jackknife, and leave-k-out analyses; (8) test residuals against every available covariate; (9) benchmark candidate models against nuisance, mass-only, and quality baselines; (10) correct for multiple comparisons across a declared test family; (11) interpret physically only after statistical validation.

Documenting discarded methods is part of the audit, because each rejection is itself a result. The following approaches were evaluated and deliberately excluded from inference, with the reason stated: tuned-protocol correlations as evidence of coupling (rejected: the tuning injects structure--residual covariance, Section 3.4); mixed radius-operator comparisons (rejected: they amplify protocol differences through geometry alone, Section 2.1); point-level BIC as primary inference (rejected: intra-galaxy points share systematics, Section 2.5); the zero-prediction cross-validation baseline as the sole benchmark (rejected: it flatters any nontrivial model; mass-only and quality baselines added, Section 3.3); LASSO variable selection as physical inference (rejected: representation-dependent within a collinear feature set, Section 3.12); point-level $\Delta\chi^{2}$ confidence regions (rejected: within-galaxy error correlation makes them underestimates, Section 3.3); and a single uniform-$\sigma$ asymmetric-drift correction as the definitive pressure-support treatment (rejected: bracketed instead between isotropic and Jeans limits, Section 3.10). Methods that survived --- the fixed protocol, galaxy-level statistics, the three-baseline predictive benchmark, hierarchical partial pooling, the canonical control specified in advance, and galaxy-level bootstrap uncertainties --- constitute the audit standard proposed in Section 5.

\section{Data Availability}

This work uses the publicly available SPARC database (\citealp{Lelli2016}). All derived data products, the analysis pipeline, SHA-256 provenance chains for the source tables, and figure-generation scripts are archived at Zenodo (concept DOI: 10.5281/zenodo.19503119; this analysis corresponds to version v20, doi:10.5281/zenodo.21856110). The archived pipeline deterministically regenerates the sample construction, residual definitions, all correlation and partial-correlation results, the threshold and inclination sweeps, the three-baseline cross-validation, the bootstrap and contour analyses, the hierarchical random-intercept fits, the scatter decomposition ($\sigma_{M0}$, $\sigma_{M3}$), and the external cross-checks. The information-criterion comparisons (the M0--M3 BIC ladder), the M2 parameter fit, the permutation and leave-k-out tests, the injection-recovery test, and the regularised-regression diagnostics derive from the original analysis notebook, archived in the same record; anchor statistics agree between the two to the stated precision, and the verification table is included in the archive.

\section{Acknowledgements}

This research used data from the SPARC database (\citealp{Lelli2016}); we thank the SPARC team for making their data publicly available. Large language models were used as writing-support and code-review tools; all numerical results and scientific claims were computed by the author using the archived pipelines and independently anchor-verified against a separate rebuild from the SPARC MRT source tables. Software: NumPy (\citealp{Harris2020}), SciPy (\citealp{Virtanen2020}), Matplotlib (\citealp{Hunter2007}), scikit-learn (\citealp{Pedregosa2011}), statsmodels (\citealp{Seabold2010}).

\bibliographystyle{plainnat}
\bibliography{references}

\begin{thebibliography}{}
\providecommand{\natexlab}[1]{#1}
\providecommand{\url}[1]{\texttt{#1}}

\bibitem[Chae et~al.(2020)]{Chae2020}
Chae K.-H., Lelli F., Desmond H., McGaugh S. S., Li P., Schombert J. M., 2020, ApJ, 904, 51

\bibitem[Desmond(2023)]{Desmond2023}
Desmond H., 2023, MNRAS, 521, 1817

\bibitem[Di Cintio et~al.(2014)]{DiCintio2014}
Di Cintio A., Brook C. B., Macci\`o A. V., Stinson G. S., Knebe A., Dutton A. A., Wadsley J., 2014, MNRAS, 437, 415

\bibitem[Duey et~al.(2025)]{Duey2025}
Duey F., Schombert J. M., McGaugh S. S., Lelli F., 2025, AJ, doi:10.3847/1538-3881/adaf21

\bibitem[Durbala et~al.(2020)]{Durbala2020}
Durbala A., Finn R. A., Crone Odekon M., Haynes M. P., Koopmann R. A., O'Donoghue A. A., 2020, AJ, 160, 271

\bibitem[Famaey \& McGaugh(2012)]{Famaey2012}
Famaey B., McGaugh S. S., 2012, Living Rev. Relativ., 15, 10

\bibitem[Harris et~al.(2020)]{Harris2020}
Harris C. R. et al., 2020, Nature, 585, 357

\bibitem[Hunter et~al.(2012)]{Hunter2012}
Hunter D. A. et al., 2012, AJ, 144, 134

\bibitem[Hunter(2007)]{Hunter2007}
Hunter J. D., 2007, Comput. Sci. Eng., 9, 90

\bibitem[Iorio et~al.(2017)]{Iorio2017}
Iorio G., Fraternali F., Nipoti C., Di Teodoro E., Read J. I., Battaglia G., 2017, MNRAS, 466, 4159

\bibitem[Kass \& Raftery(1995)]{Kass1995}
Kass R. E., Raftery A. E., 1995, J. Am. Stat. Assoc., 90, 773

\bibitem[Katz et~al.(2017)]{Katz2017}
Katz H., Lelli F., McGaugh S. S., Di Cintio A., Brook C. B., Schombert J. M., 2017, MNRAS, 466, 1648

\bibitem[Keller \& Wadsley(2016)]{Keller2016}
Keller B. W., Wadsley J. W., 2016, ApJ, 835, L17

\bibitem[Lelli et~al.(2016)]{Lelli2016}
Lelli F., McGaugh S. S., Schombert J. M., 2016, AJ, 152, 157

\bibitem[Lelli et~al.(2017)]{Lelli2017}
Lelli F., McGaugh S. S., Schombert J. M., Pawlowski M. S., 2017, ApJ, 836, 152

\bibitem[Li et~al.(2018)]{Li2018}
Li P., Lelli F., McGaugh S. S., Schombert J. M., 2018, A\&A, 615, A3

\bibitem[Marra et~al.(2020)]{Marra2020}
Marra V., Rodrigues D. C., de Oliveira A. O. F., 2020, MNRAS, 494, 2875

\bibitem[McGaugh et~al.(2016)]{McGaugh2016}
McGaugh S. S., Lelli F., Schombert J. M., 2016, Phys. Rev. Lett., 117, 201101

\bibitem[Meidt et~al.(2014)]{Meidt2014}
Meidt S. E. et al., 2014, ApJ, 788, 144

\bibitem[Milgrom(1983)]{Milgrom1983}
Milgrom M., 1983, ApJ, 270, 365

\bibitem[Mistele et~al.(2024)]{Mistele2024}
Mistele T., McGaugh S. S., Lelli F., Schombert J. M., Li P., 2024, ApJ, 969, 27

\bibitem[Oh et~al.(2015)]{Oh2015}
Oh S.-H. et al., 2015, AJ, 149, 180

\bibitem[Oman et~al.(2019)]{Oman2019}
Oman K. A. et al., 2019, MNRAS, 482, 821

\bibitem[Pedregosa et~al.(2011)]{Pedregosa2011}
Pedregosa F. et al., 2011, J. Mach. Learn. Res., 12, 2825

\bibitem[Pontzen \& Governato(2012)]{Pontzen2012}
Pontzen A., Governato F., 2012, MNRAS, 421, 3464

\bibitem[Read et~al.(2019)]{Read2019}
Read J. I., Walker M. G., Steger P., 2019, MNRAS, 484, 1401

\bibitem[Rodrigues et~al.(2018)]{Rodrigues2018}
Rodrigues D. C., Marra V., del Popolo A., Davari Z., 2018, Nature Astron., 2, 668

\bibitem[Schombert et~al.(2019)]{Schombert2019}
Schombert J. M., McGaugh S. S., Lelli F., 2019, MNRAS, 483, 1496

\bibitem[Seabold \& Perktold(2010)]{Seabold2010}
Seabold S., Perktold J., 2010, Proc. 9th Python in Science Conf., 92

\bibitem[Var\u{a}\c{s}teanu et~al.(2025)]{Varasteanu2025}
Var\u{a}\c{s}teanu D. et al., 2025, MNRAS, 541, 2366

\bibitem[Virtanen et~al.(2020)]{Virtanen2020}
Virtanen P. et al., 2020, Nat. Methods, 17, 261

\bibitem[de Blok \& McGaugh(1997)]{deBlok1997}
de Blok W. J. G., McGaugh S. S., 1997, MNRAS, 290, 533

\end{thebibliography}
\end{document}